\documentclass[a4paper, twocolumn]{article}
\usepackage{graphicx}
\usepackage{amsmath,amssymb}
\usepackage{verbatim}
\usepackage{color}
\usepackage{textcomp}

\date{}

\title{Measurement of the complete interaction force curve at the nanoscale}

\author{Mario S. Rodrigues$^{1,2}$, Luca Costa$^{2,3}$, Jo\"{e}l Chevrier$^{2,3,4}$, Fabio Comin$^2$ \\
\\
\small 1- Departamento de F\'{\i}sica, Universidade de Lisboa, Campo Grande 1749-016 Lisboa, Portugal\\
\small \texttt{mmrodrigues@fc.ul.pt}\\
\small 2- European Synchrotron Radiation Facility , 6 rue Jules Horowitz BP 220, 38043 Grenoble Cedex, France \\
\small 3- Universit\'{e} Joseph Fourier BP 53, 38041 Grenoble Cedex 9, France\\
\small 4- Institut N\'{e}el CNRS BP 166, 38042 Grenoble Cedex 9, France \\
\begin{minipage}{.85\textwidth} 
\vspace{12pt}
\small
The force between two interacting particles as a function 
of distance is one of the most fundamental curves in science. 
In this regard, Atomic Force Microscopy (AFM) represents the most powerful tool in nanoscience 
but with severe limits when it is to probe attractive interactions with high sensitivity. 
The Force Feedback Microscope (FFM) described here, removes from AFM 
the well known jump to contact problem that precludes the complete exploration 
of the interaction curve and the study of associated energy exchanges. 
The FFM makes it possible to explore tip-surface interactions in the entire range 
of distances with a sensitivity better than 1 pN. 
FFM stands out as a radical change in AFM control paradigms. 
With a surprisingly simple arrangement it is possible to provide the AFM tip 
with the right counterforce to keep it fixed at any time. 
The counterforce is consequently equal to the tip-sample force. 
The force, force gradient and damping are simultaneously measured independently of the tip position. 
This permits the measurement of energy transfer in thermodynamic transformations. 
Here we show some FFM measurement examples of the complete interaction force curve 
and in particular that the FFM can follow the nucleation of a water bridge 
by measuring the capillary attractive force at all distances, 
without jump to contact despite the large attractive capillary force. 
Real time combination of the measured parameters will lead to new imaging 
modalities with chemical contrast in different environments.
\end{minipage}
}

\begin{document}

\maketitle

As stated by Feynman in his lectures \cite{1}, 
the force versus the distance between two interacting 
atoms is of the utmost importance in science 
being at the basis of our understanding of interactions between two objects. 
From the sharp repulsive regime felt at very short distances, 
a daily manifestation of the Pauli repulsion principle, 
the interaction extends for many tens or hundreds of nanometers in a (usually) long attractive regime.
In this regime different sources (electrostatic, magnetic, chemical, capillary, van der Waals) 
at different scales intervene, and the quantitative measurement of the interaction 
over the entire span is essential for the understanding of the underlying mechanisms. 
This is well reflected by the wealth of AFM experimental activity in different environments 
and thermodynamical conditions covering physics, mechanics, biology, chemistry and soft condensed matter. 
However, despite the great successes obtained by AFM \cite{2, 3, 4, 5} 
and Surface Force Apparatus (SFA) \cite{6, 7}, up to now there is no instrument 
that can systematically and directly provide in real time the full force curve 
at nanoscale 
and at the pN level
as a real 
unambiguous 
experimental result. 
The SFA integrates over microareas, while in AFM, most of the time, 
an uncontrolled and irreversible dive of the nanotip onto the surface 
prevents direct and immediate access to the interaction in a large portion of the attractive regime. 
This \textit{jump to contact} intervenes as soon as the force gradient 
overcomes the stiffness of soft AFM microlevers used for high sensitivity probing.
Despite previous efforts to overcome this limitation (see for ex.: \cite{b8a,b8b,b8c}),
it still remains intrinsic to the AFM paradigm. 
It 
is rather unfortunate because if the repulsive interaction is essential to 
locate atoms and to obtain surface topography, 
the chemical and physical specificities of the surface of the nano-objects deposited there, 
can only be characterized by measuring the non contact attractive interaction curve. 
Moreover, Ritort et al. \cite{9} have experimentally shown that 
thermodynamic quantities such as the folding free energy of RNA 
can be extracted from controlled measurements of non equilibrium cycles using the 
Crooks fluctuation theorem \cite{10}. 
They used optical tweezers to continuously measure position and force during folding and unfolding of 
RNA hairpins. Because FFM provides full control of tip position during force measurement, 
this opens the possibility to extend this strategy to forces ranging from piconewton to above a nanonewton.

The Force Feedback Microscope described here measures 
the interaction force in its entire range. 
An example is given in Fig. 1 where the complete force curve as measured 
by the FFM is compared with a standard AFM approach. 
The system is the prototypical mica surface in deionized water explored with a silicon nitride tip. 
In liquids, charges between tip and surface may induce a non contact repulsion. 
The combination of this electrostatic interaction with van der Waals attraction 
leads to a characteristic behavior visible in Fig. 1 and described by the DLVO theory 
\cite{7,11}.

\begin{figure}[ht]
\centering \includegraphics[width=.9\linewidth]{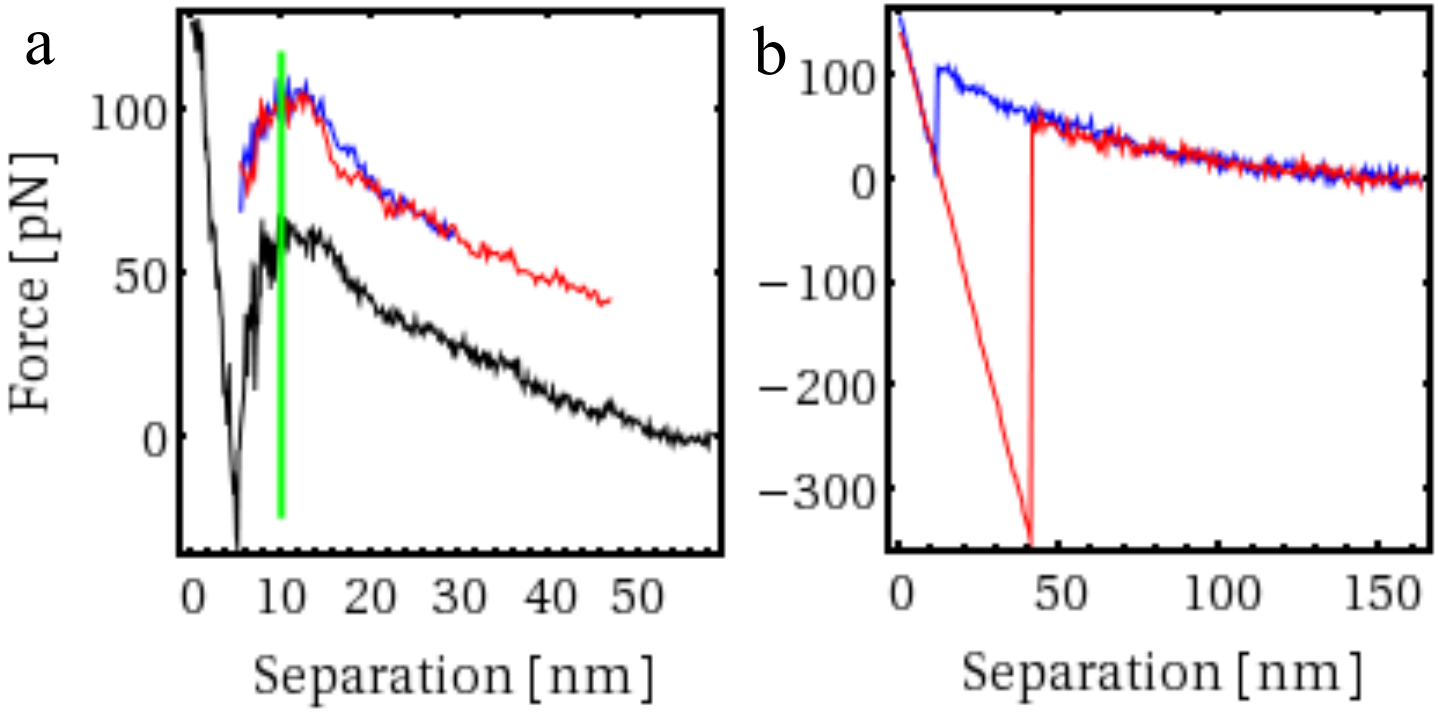}
\vspace{-8pt} \caption{ \small Force curves measured with silicon 
nitride tips on a cleaved mica surface immersed in 
deionized water. 
\textbf{a}, the FFM is used to measure the full force curve 
(black curve) with a lever of stiffness $k= 0.014$ N/m.
The vertical green bar marks the location of the 
jump to contact if the FFM protocol would be deactivated 
and a regular AFM approach curve would be performed.
The blue and red curves (shifted upward for clarity) 
show that it is possible to track reversibly back and forth 
and with no hysteresis any portion of the attractive part of the interaction.
\textbf{b}, AFM conventional static mode with a lever of stiffness 
$k= 0.028$ N/m. 
As the tip approaches the surface, 
the uncontrolled jump to contact occurs when the gradient 
of the force exerted by the surface on the tip exceeds $k$. 
The characteristic hysteresis curve is well visible.
}
\label{fig.1}
\end{figure}

AFM operations are based on the static contact mode 
and on the amplitude modulation mode (AM-mode) \cite{12} 
with all its associated variants such as Frequency Modulation mode \cite{13}. 
The dynamical modes probe the tip-surface interaction by quantifying 
the perturbation that this interaction induces on the lever oscillations as 
it is excited at resonance. 
In the linear regime, i.e. when the oscillation amplitudes are kept small, 
the force gradient exerted on the tip by the surface determines a shift in resonance frequency. 
The oscillating modes can be easily activated either on the attractive or on 
the repulsive portion of the interaction with no jump to contact. 
However they face two contradictory constrains: on one side, 
the amplitude of vibration has to be kept small to avoid non linearities 
induced by the curvature of the force curve and on the other, 
the amplitude of vibration should be large enough to have a high sensitivity 
in detecting the force gradient and to avoid jump to contact when the lever is soft. 
All this leads to clear difficulties in the use of oscillating modes 
for measuring directly and in real time the force gradient $\triangledown F$ and the 
damping $\gamma$ with high resolution. 
Furthermore, the force sensitivity is essentially controlled by the $Q/k$ ratio where $Q$ and $k$
are respectively the quality factor and stiffness of the cantilever. 
This ratio acts as the gain of a mechanical amplifier. 
In liquids, the large decrease of $Q$ down to values close to one 
essentially suppresses the key advantage of working at resonance. 
Many research programs \cite{14} are nowadays pursued to overcome these difficulties
that affect AFM oscillating modes since their inception. 

Here, the force is obtained by measuring the counterforce 
that must be applied to keep the tip position $X_{tip}$ constant. 
A feedback loop, shown in blue in Fig. 2 generates an external 
force that offsets the interaction force on the tip:
$F_{surface/tip}(t) = - k (X_{ps}(t) - X_{tip})$, 
where $X_{ps}(t)$ is the position of the clamped extremity of the microlever 
on which a piezoelement acts. 
The knowledge of $X_{ps}(t)$ gives the force $F_{surface/tip}$ whatever 
the distance between the surface and the tip. 
As the jump to contact is prevented, 
AFM levers with very low stiffness $k$ can be used leading to increased force sensitivity.

\begin{figure}[ht]
\includegraphics[width=1\linewidth]{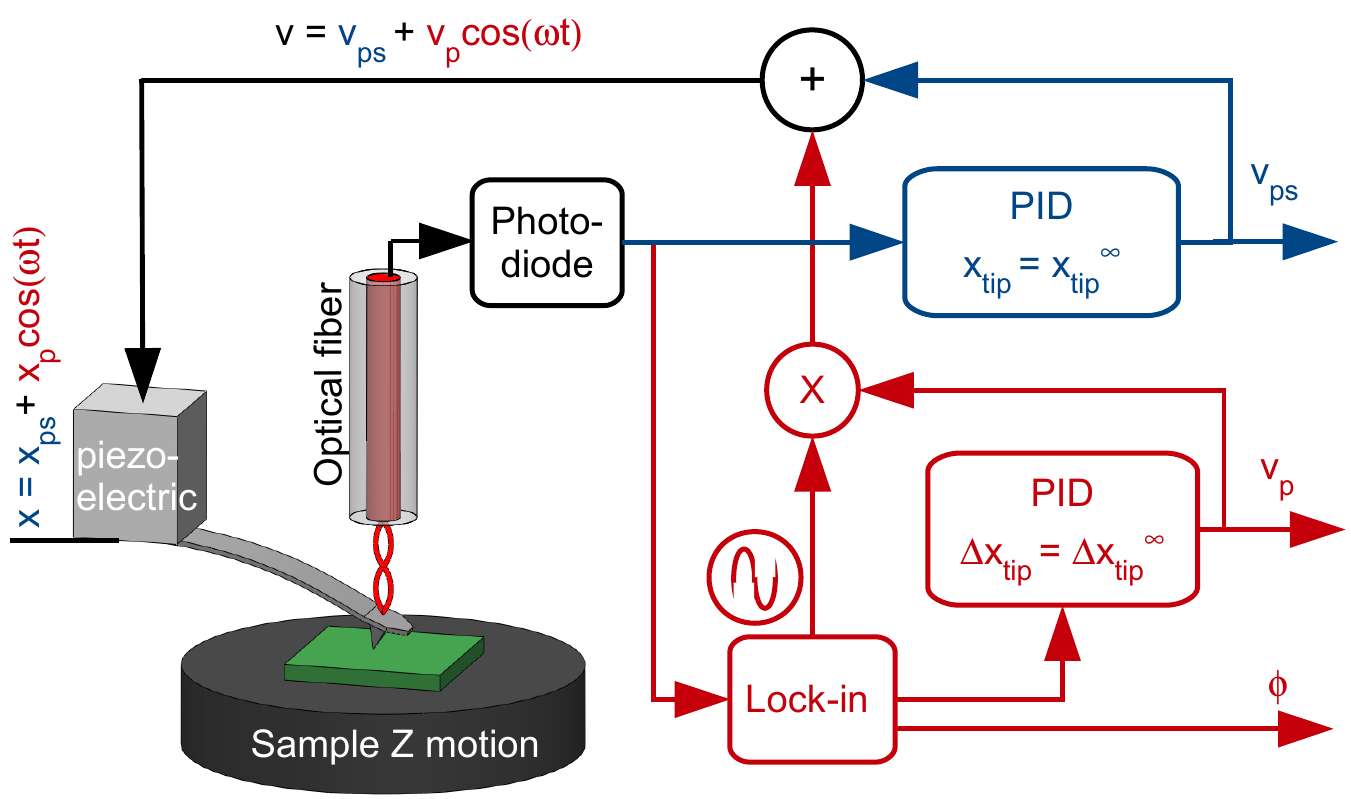}
\vspace{-8pt} \caption{\small Block diagram of the FFM operation. Static mode (blue loop): 
$X_{ps}$ via the lever stiffness $k$ gives directly the static forces acting on the tip.
Dynamic mode (red loop): the tip oscillation $\Delta X_{tip}$ is kept constant. 
The voltage $v_p(d,t)$  and the phase $\phi (d)$ applied to the piezoelement 
are measured quantities from which $\triangledown F(d)$ and $\gamma (d)$ are determined (see text).
}
\label{fig.2}
\end{figure}

Imaging can be done by changing the sample position so that the force 
between tip and surface is kept constant. 
The elements constituting the FFM are shown in Fig. 2.  
Rather than using the lever deflection, 
the precise position of the tip is measured by using a cleaved optical fiber 
positioned at the back of the lever \cite{15}. 
The distance between the lever and the fiber end is around 10$\mu$m, 
constituting a Fabry Perot cavity with a sensitivity close to 
1 pm$\mathrm{Hz^{-1/2}}$ \cite{16,17}.
The cavity is used to calibrate the piezoelement response. 
Calibration of the lever stiffness $k$ is done using the thermal method and results in a 
10\% uncertainty \cite{18,19}.
Since displacements of the piezoelement end position $X_{ps}$ as low as 10 pm can be detected, 
the use of a cantilever with 0.014 N/m stiffness (Fig. 1)
brings the theoretical limit in sensitivity below the piconewton regime.

To be noted that in an FFM experiment, an abrupt change in the interaction force 
does not lead to any change in distance between the tip and the surface. 
This transitory in force can be explored at ease by slowly 
and continuously varying the tip surface distance. 
A typical occurrence is when a capillary water bridge nucleates between a nanotip and the surface \cite{20}. 
The measure of this sudden and strong force variation is presented in Fig. 3 
together with the related increase of the attractive force as the tip is slowly moved 
toward the surface shortening the capillary bridge.

\begin{figure}[ht]
\centering \includegraphics[width=1\linewidth]{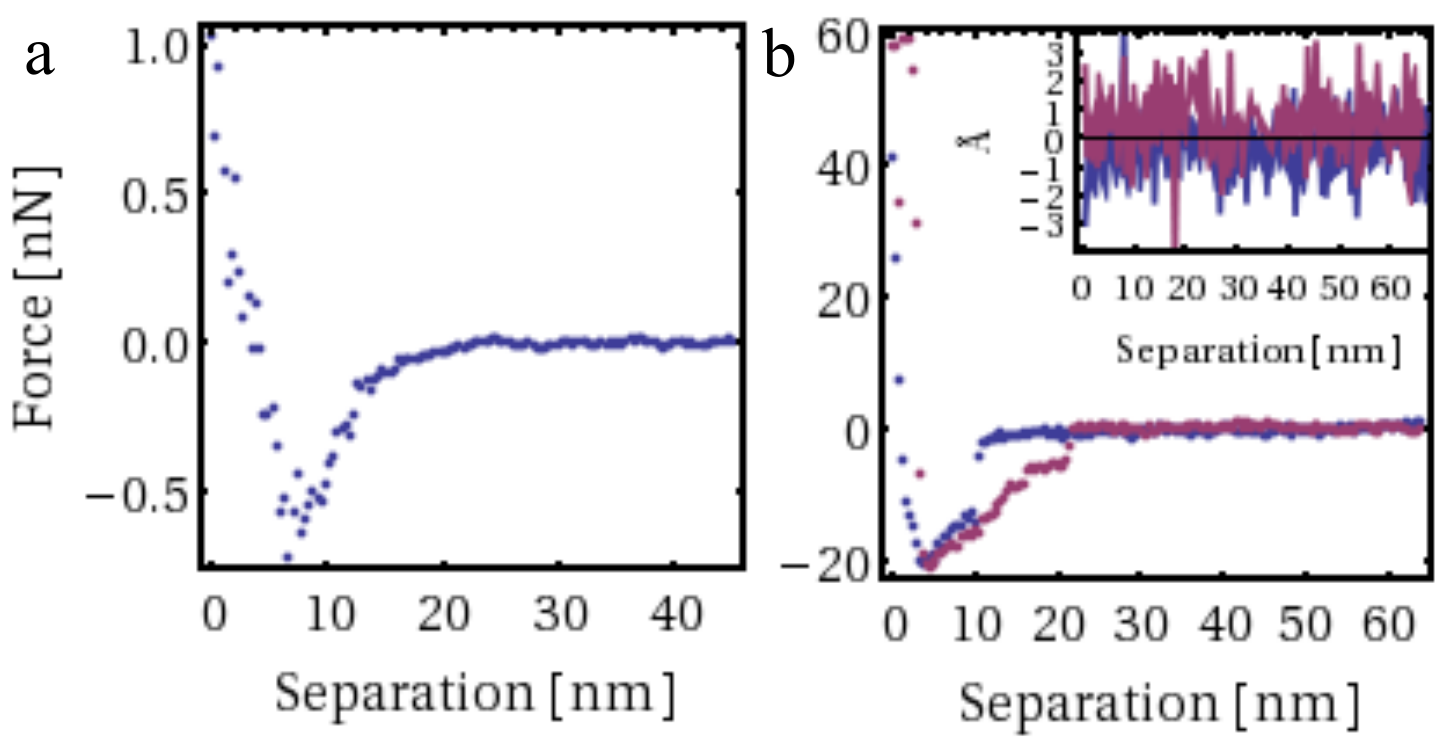}
\vspace{-8pt} \caption{\small Force curves measured by FFM in air on a hydrophobic 
cleaved graphite surface (\textbf{a}) and on hydrophilic silicon native 
oxide surface (\textbf{b}) using a silicon tip attached to a lever of stiffness 
$k$ = 0.35 N/m (right) and $k$=0.15 N/m (left). 
\textbf{a}, FFM reversibly measures the tip surface interaction with no jump to contact.
\textbf{b}, when the tip surface distance is around 10nm, 
a capillary bridge suddenly forms and the force abruptly increases to about 10nN. 
When the tip is further approached, the force keeps increasing up to 20nN. 
On the return path the hysteresis is well visible. 
The hysteresis is intrinsic to the capillary bridge properties 
and not related to the apparatus response. 
The inset shows the position of the tip as measured during the experiment.}
\label{fig.3}
\end{figure}

\noindent The FFM is a new tool to measure forces associated to fluid behaviors at nanoscale. 
The possibility of measuring non-equilibrium transformations to determine 
equilibrium thermodynamic quantities \cite{9} should find here another application even when 
the intervening forces are larger than 1nN, 
exceeding the capabilities of optical tweezers.
The hysteresis loop presented in Fig. 3 is determined by the intrinsic character 
of the capillary bridge condensation and rupture. 
Repeated measurements of hysteresis cycles under different external conditions 
such as hygrometry level, 
should lead to the determination of the associated free energy after data processing 
based on a formal treatment similar to the one used by Ritort et al., 
opening the possibility of a detailed and quantitative study of the capillary bridge between tip and surface.

With the first feedback loop active, 
a second feedback loop is activated to ensure the tip oscillates at a chosen frequency $ω$ 
(not necessarily 
close to the
the resonance frequency) and 
with a constant oscillation amplitude $\Delta X_{tip}$. 
As the tip oscillates close to the surface, $\triangledown F$ and $\gamma$ are simultaneously 
determined from the detailed knowledge of the oscillating voltage and phase applied to the piezoelement 
to keep the amplitude constant. 
Here, the force applied by the lever to the tip, $F_{lever/tip} = k(X_{ps} (t) - X_{tip})$, 
keeps the average position of the tip, $X_{tip}$ constant and prevents the jump to contact. 
In addition, an excitation at a selected frequency $ω$ is controlled by a second feedback loop, 
in order to keep the oscillation amplitude $\Delta X_{tip}$ 
constant in all the interaction range. 
The frequency $ω$ can be chosen arbitrarily. 
In the example shown here the frequency is 1.1 kHz. 
The oscillation amplitude $\Delta X_{tip}$ is kept reasonably large but below values 
characterizing features of the force curves, 
which are typically of some tens of pN.

As shown in Fig. 2, in the FFM dynamic mode, 
the measured experimental values are the voltage amplitude $v_p(d)$ 
($v_p^\infty $ when $d$ is made very large) 
needed to keep the tip oscillation amplitude constant and the phase $\phi(d)$ 
($\phi^\infty $ when $d$ is made very large) between the lever oscillation and the excitation. 
The force gradient $\triangledown F(d)$ and the damping $\gamma$ 
are derived directly from $v_p(d)$, $v_p^\infty $, $\phi(d)$ and $\phi^\infty $.
The force gradient can be written 
using the normalized lever voltage excitation $v_n(d) = v_p(d)/v_p^\infty $: 

\begin{equation}
 \triangledown F(d) =  a [\cos(\phi^\infty ) - v_n(d) \cos(\phi(d))]
\end{equation}

\noindent The damping can be written using a similar and closely related expression:

\begin{equation}
 \gamma (d) = \frac{a}{\omega} [-\sin(\phi^\infty ) + v_n(d) \sin(\phi(d))]
\end{equation}

\noindent Parameters $a$ and $\phi^\infty $ are linked to the physical probe employed, 
to the FFM set up and to the experimental environment (liquid, vacuum, gas,...). 
They are determined by the dynamic response function of the instrument and are therefore 
frequency dependent but not dependent on the tip sample interaction. 
The expressions for $a$ and $\phi^\infty $ are straightforward to deduce considering 
the cantilever as a mass-spring system \cite{a}.
{
To note that these formulas include the information of the lever motion as well as motion of
the base of the cantilever. 
Here the total motion of the tip (deflection + base motion) is measured trough the optical fibber,
not just the cantilever deflection. This avoids difficulties as those mentioned in ref. \cite{issues}.
The constant $a$, whose units are N/m, is the equivalent of a dynamic stiffness while 
$\phi^\infty $ is related to the loss of energy to the surrounding media. 
These constants can be measured in the real experimental environment 
to define the instrument calibration when FFM dynamic mode is used. 
This is very similar to the determination of $k$ and $Q$ to which $a$ and $\phi^\infty $ are closely related, 
in more conventional AFM modes. 
No further a posteriori treatment is needed to access the details of the tip surface interaction \cite{remark}. 
In FFM dynamic mode, images can be obtained using regular AFM strategies, i.e. 
imaging can be done rastering the sample keeping the force gradient $\triangledown F$ constant.

\begin{figure}[ht]
\centering \includegraphics[width=1\linewidth]{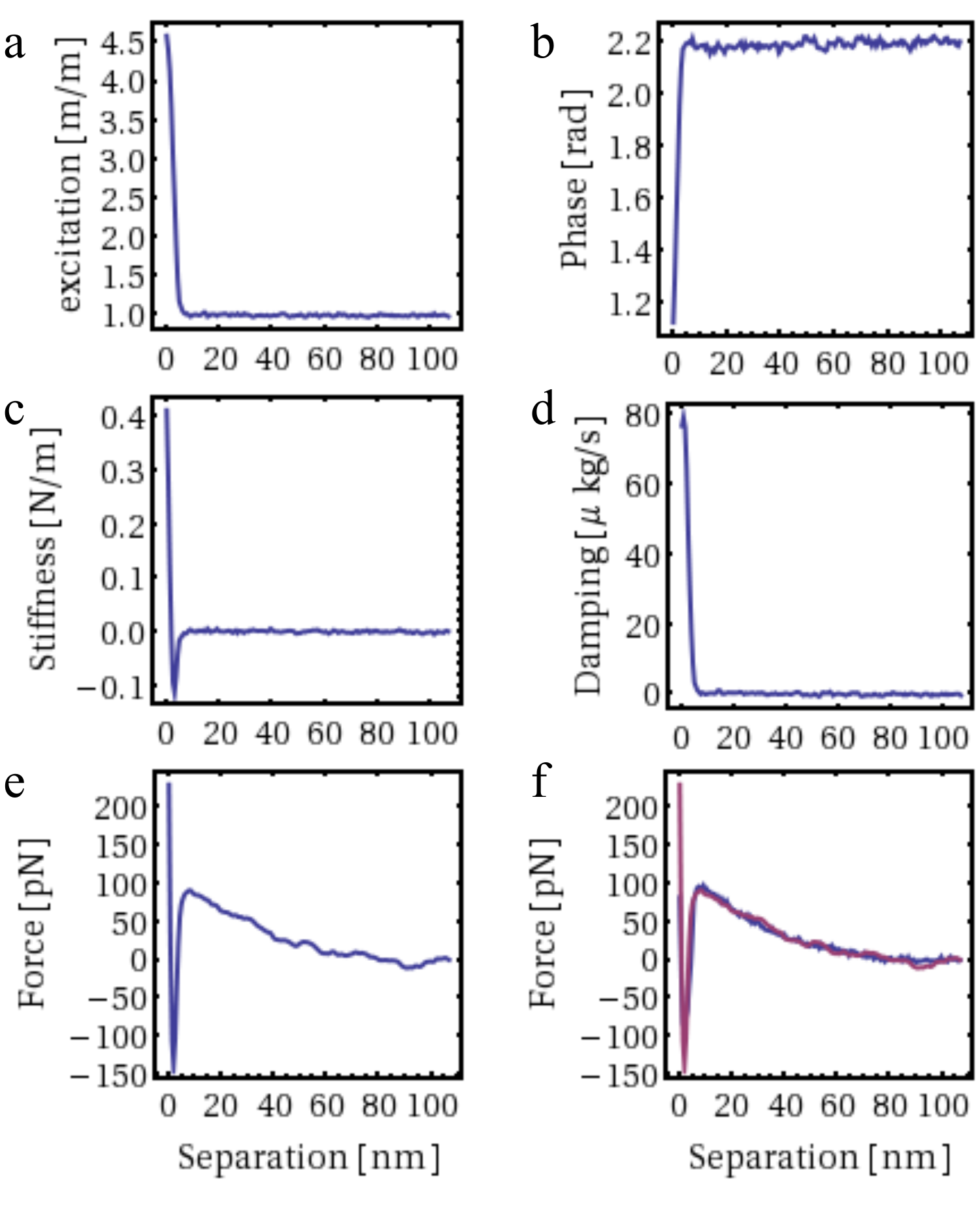}
\vspace{-8pt} \caption{ \small Force curves. 
Experimental conditions: 
Classical silicon 
nitride tip on a lever of stiffness $k=$ 0,015 N/m. 
The tip interacts with the surface of freshly cleaved mica. 
The measurement is performed in deionized water at $f = $ 1125 Hz. 
The tip oscillation $\Delta X_{tip}$ is 0.2nm. 
\textbf{a}, normalized piezoelectric excitation $v_n(d) = v_p(d) / v_p^\infty$ 
where $v_p(d)$ is the voltage applied to the piezoelement to 
keep the tip oscillation amplitude constant and $v_p^\infty = v_p(d)$  as $d$ is very large;
\textbf{b}, $\phi(d)$ is the phase between $v_p(t,d)$ and the tip oscillation. 
\textbf{c}, $\triangledown F(d)$ determined using equation 1 applied to data \textbf{a} and \textbf{b}; 
\textbf{d} damping $\gamma (d)$ determined using equation 2; 
\textbf{e} result of the numerical integration of $\triangledown F(d)$ i.e. of the curve \textbf{c}; 
in \textbf{f} the numerical integration of $\triangledown F(d)$ is compared to the measured force.
}
\label{fig.4}
\end{figure}

The simultaneous and independent measurement of the force and force gradient 
as a function of distance confers another original and key advantage to FFM 
since static and dynamic information can be extracted in parallel. 
As an example, Fig. 4 shows a complete sequence of measurements and real time data analysis 
leading to the determination of force approach curve using the two methods, 
the static and the dynamic one. Panel f in Fig. 4 contains the direct measurement of the force curve 
$F(d)$ using the static mode. A second method to obtain $F(d)$ is based on the numerical 
integration of the force gradient $\triangledown F(d)$ determined using the measurements of $v_n(d)$ 
and $\phi_(d)$ shown in Fig. 4a and 4b respectively. 
The constants $a$ and $\phi^\infty $ are determined just once from best fitting a measured static force curve. 
For any given fixed instrumental configuration the parameters $a$ and $\phi^\infty $ can be treated 
as calibration constants. 
In the specific case their values are $a = 0.051 $N/m and $φ^\infty  = -54$ Deg. 
Comparison of the integrated force gradient (Fig. 4e) and of measured force (Fig. 4f) 
shows a strong reduction in noise from about 10 pN in the broadband direct static force measurement, 
to 1 pN in the numerically integrated force gradient that has been measured using a narrow 
lock-in bandwidth around the excitation frequency $\omega$. 
Fig. 4 characterizes a measurement strategy that besides determining $\triangledown F(d)$ 
the force gradient and the damping $\gamma$, provide a complete measurement of the force curve 
$F(d)$ with piconewton sensitivity from the combined and simultaneous use of the two FFM modes. 
We believe that, in the large family of Scanning Probe Microscopy, 
among the different AFM modes, the FFM represents a new approach when it is to 
measure simply and directly forces at nanoscale and at the piconewton level. 
The FFM enables a user to have the full control of the tip surface interaction 
(no sudden jump to contact) and to describe its evolution versus tip surface distance in any environment. 
Vacuum conditions or operation in liquids do not affect the principle and measurements can be 
carried out easily also on those environments. 
The force, either attractive or repulsive, the force gradient, 
and the damping are all directly measured on line. 
For example, FFM opens the possibility to routinely measure images in non contact static mode. 
FFM provides a new tool to explore challenging questions such as properties of fluids at nanoscale.

\noindent \textbf{Acknowledgments}

\noindent We thank E. Charlaix and G. Cappello for discussions. 
This work was performed at the Surface Science Laboratory of the ESRF in Grenoble. 
M.S.R. acknowledges support from the project ANR-09-NANO-042-02 PianHo.


\begin{thebibliography}{30}
\bibitem{a}   Considering a mass spring system driven by a motion of the base to which the spring is attached
then the constants take the form $a= \sqrt{(k-m \omega^2)^2+\gamma_m^2 \omega^2}$
and $\phi^\infty= \arctan[\gamma_m \omega/(\omega^2-\omega_0^2)]$ with $\gamma_m=\gamma/m$.
\bibitem{remark} Here, it is not assumed the lever to behave like a simple mass-spring system 
and hence the constants $a$ and $\phi^\infty$ are measured.
\end{thebibliography}
\end{document}